# LYMAN $\alpha$ ABSORPTION AND TIDAL DEBRIS

SIMON L. MORRIS

*Dominion Astrophysical Observatory, National Research Council, 5071 West Saanich Road, Victoria, B.C., V8X 4M6, Canada*

## ABSTRACT

The origin and evolution of structure in the Universe is one of the major questions occupying astronomers today. An understanding of the Ly$\alpha$ absorbers seen in QSO spectra is an important part of this program since such absorbers can be traced back to very high redshifts. Their mere existence places constraints on the physical state of the intergalactic medium. The discovery of Ly$\alpha$ absorbers at low redshift allows us to estimate for the first time what fraction of low redshift Ly$\alpha$ absorbers are (i) randomly distributed, (ii) distributed like galaxies but not physically associated with luminous objects, (iii) actually part of the halos of luminous galaxies, or (iv) tidal tails within galaxy groups. Results from the sightline to the QSO 3C273 suggest that the majority of the absorbers are not associated with galaxies, but that there is a significant subset that are. The absorbers associated with galaxies may be produced in enormous gaseous disks surrounding normal spiral galaxies, or may be tidal material bound up in small groups of galaxies

## 1. HST UV Spectroscopy

Prior to 1991, study of Ly$\alpha$ absorbers was restricted to redshifts greater than 1.6, where the lines were shifted into the visible region of the spectrum. Recently, however, it has been discovered that Ly$\alpha$ absorbers can be detected at very low redshift by the UV spectrographs on HST, with the best studied example being the sightline to 3C273[11,1]. These results indicate the presence of at least 11, and possibly as many as 16, Lyman $\alpha$ absorbers distributed over redshifts from 0 to 0.158. A much larger sample of Ly$\alpha$ absorption lines spread over many QSO lines of sight (LOS) is being collected as part of the HST Key Project (KP) on QSO absorption lines[3]. These recent discoveries offer an exciting opportunity to investigate directly the kinds of galaxies and clusters, if any, with which the Ly$\alpha$ absorbers are associated at low redshifts.

The number of absorbers per redshift interval was seen from ground based spectroscopy to be dropping very rapidly from high redshift towards z=1.6. This rapid decline seems to level off at low redshifts, although the exact form of the low redshift dN/dz is not yet known. This behaviour could indicate either that there are 2 different populations of absorbers - one dominating at high redshift, that has vanished by the present day, leaving a different residual 'constant' population, or alternatively a single absorber population whose dN/dz changes slope at a redshift around 1.5. Both of these pictures are currently viable.

One of the striking properties of the Ly$\alpha$ 'forest' at *high* redshift is the lack of significant absorber-absorber clustering, in contrast to the higher column density absorption systems (which also show detectable C IV). This is generally taken to imply that the majority of these absorbers are not associated with galaxies.

## 2. Galaxy Redshift Surveys

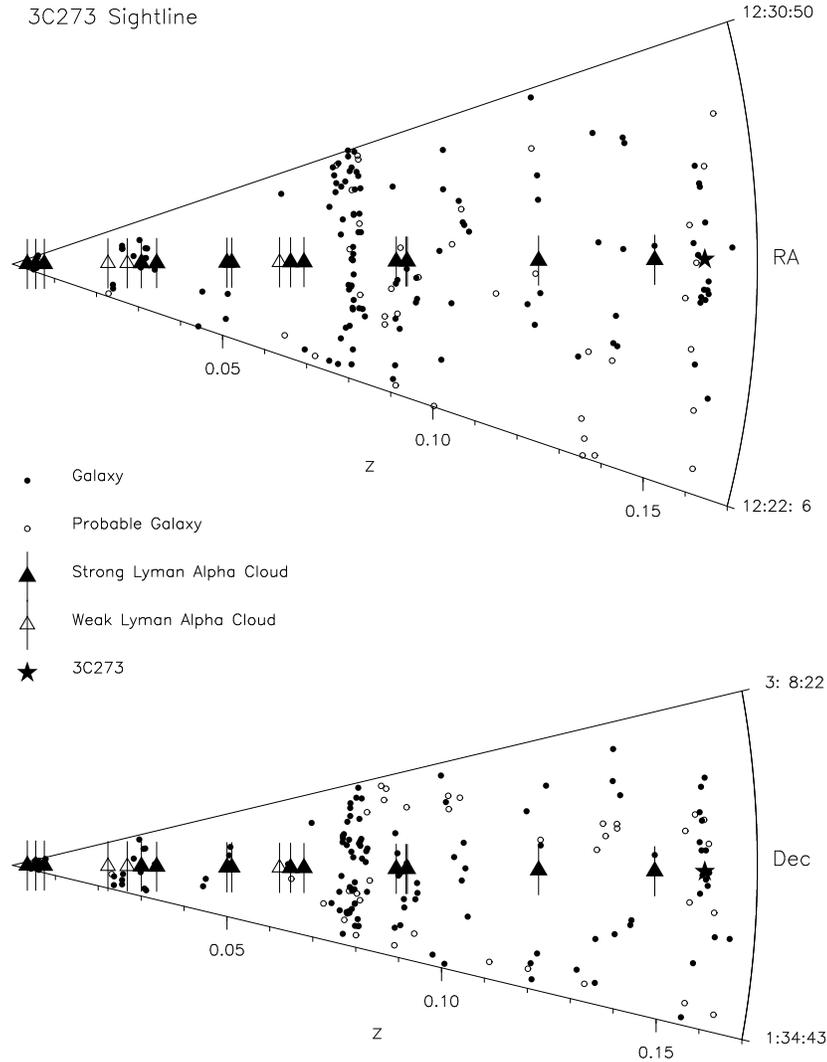

Fig. 1. Pie-diagrams for the Galaxies observed in the 3C273 LOS with the LCO fiber system[12]. Angles have been exaggerated to prevent overcrowding of the symbols. Please note that this results in a highly distorted plot with initially spherical structures (such as the 3C273 cluster of galaxies) appearing elongated transverse to the line of sight. Note also that the star marking the position of 3C273, while readily visible in the projection in RA, is partially obscured by a clump of galaxies in the projection in Dec

An obvious next step is to combine galaxy redshift catalogs around the QSO

LOS with UV spectroscopic data in order to measure the absorber-galaxy correlation function at low redshift. Morris et al.[12] presented data on a sample of 176 galaxies within a radius of 45 arcmin of the 3C273 LOS, complete to B~19. Fig. 1 shows a pie diagram of these data.

In a complementary study, Lanzetta et al.[9] publish preliminary results for a redshift survey of 6 QSO LOS containing 26 Ly$\alpha$ absorbers. The goal of their survey is to observe all galaxies brighter than r=21.5 within a radius of 1.3 arcmin of the QSO LOS. They are presently 37% complete, with a sample of 46 galaxies. Studies of individual QSO LOS have also been published[2,16].

## 3. The Absorber-Galaxy Correlation Function

The low redshift Ly$\alpha$ absorbers in the 3C273 LOS do have a significant correlation with galaxies, but the absorber-galaxy correlation function is much weaker on large (10 Mpc) scales than the galaxy-galaxy correlation function. There is also a marginally significant result that the low redshift absorbers avoid regions of particularly high galaxy density, such as that seen in Fig. 1 at z=0.08. Mo and Morris[10] have set limits on the makeup of the absorbers, considering three simple model populations: (i) randomly distributed absorbers, (ii) absorbers with the same distribution as galaxies, but not physically part of detected galaxies, and (iii) absorbers which are part of detected galaxies (i.e. galaxy halos). For this test, the amplitude of the absorber-galaxy correlation function is estimated by counting the number of galaxies within a fixed radius of each absorber. This can be compared with counts produced using Monte Carlo samples of randomly positioned absorbers drawn from the absorber selection function (the 'random' population), or by choosing galaxies from the survey as centers for the counts (the 'galaxy-like' or 'halo' populltations)[10]. A mixture of 75% 'random' and 25% 'halo' absorbers give the best match to the data, although a range of other mixes are permitted, as shown in Fig. 2

The majority of the absorbers have to be uncorrelated with galaxies in order to produce the weak large scale correlation. The observed small-scale correlation with galaxies can be produced by a relatively small admixture of halo absorbers (as little as 10%), although the data is also consistent with galaxy halos producing up to 30% of the observed absorption lines. If none of the absorption is produced by such halos, i.e. if there are no absorbers physically associated with a galaxy in our sample with measured redshifts, then the observed correlation is marginally consistent with a 50:50 mix of random and galaxy-like absorbers.

Lanzetta et al.[9] claim that a much higher fraction of the Ly$\alpha$ absorbers in their QSO LOS are produced by material in the halos of galaxies (68% ±18%) than Mo and Morris found. This discrepancy is not yet statistically significant, but even if confirmed could be due to the different ranges of absorber column density in the two studies. While the Lanzetta et al. survey covers a considerably larger total pathlength, the HST FOS data used did not probe to as low a column density as the GHRS data.

Another result that deserves further study is the correlation claimed by Lanzetta et al.[9] between Lyα absorber equivalent width (EW) and impact parameter to the nearest galaxy. Such a correlation must clearly be present at some level given that at impact parameters of <10-20 kpc one would expect a damped Lyα absorber, while at large impact parameters there should be no absorption. However, the details of the falloff in EW with impact parameter could determine which of the models described in the next section are correct. Unfortunately, the large number of upper limits in the Lanzetta et al.[9] data set make such analysis uncertain at present.

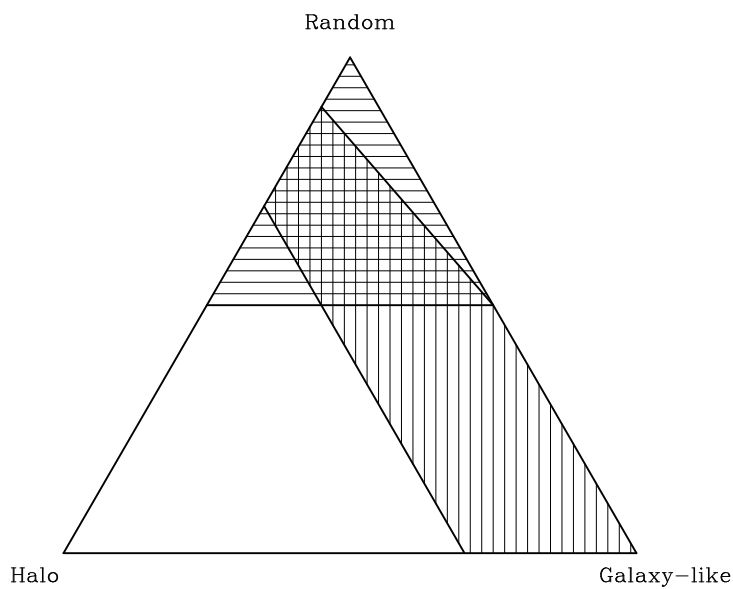

Fig. 2. Schematic diagram showing the permitted combinations of the 3 model populations[10]. The vertices of the triangle represent models with a single type of absorber (random, galaxy-like or halo). Moving along the sides of the triangle represents different admixtures of 2 different types of absorber, while the inner regions represent mixtures of all 3 types of model. Marked along the edges are the permitted ranges found by Monte-Carlo tests. For simplicity we have just drawn straight lines between these bounds to roughly illustrate the areas with 3 absorber populations which are permitted. Horizontal shading shows the region permitted by the absorber-galaxy correlation on large (10 Mpc) scales, while vertical shading shows the region permitted by the absorber-galaxy correlation on small (0.5 Mpc) scales. It can be seen that there is a substantial region permitted by both.

## 4. Models

The weak or absent absorber-absorber correlation at high redshifts motivated a large number of models for the absorbers as isolated primordial clouds of hydrogen, unassociated with galaxies. These can be crudely separated into two groups - those using pressure to confine the clouds (see for a recent example Petijean et al.[14] and references therein), and those using gravity (see for a recent example Miralda-Escude and Rees[8] and references therein). Composite models combining both of these mechanisms have also been explored for example by Charlton et al.[4].

The preliminary results at low redshift suggesting that there may be a significant absorber-galaxy correlation has produced a similar number of models attempting to place the absorbers in or near galaxy halos. Maloney[7] and Hoffman et al.[5] suggest that most if not all of the low redshift absorbers could be produced in the very extended HI disks of spiral galaxies. Due to their high ionisation (photoionised by the UV background radiation) such disks would be invisible to HI 21 cm observations. Salpeter[15] takes this idea a step further in speculating that the absorbers that are not associated with bright galaxies could be the outer parts of galaxies which formed in voids and blew all the gas out of their inner (star forming) regions, leaving a rapidly fading and soon invisible stellar disk, and large residual HI disks.

Morris and van den Bergh[13] point out that the large cross sections required for a 'halo' model to work ($\geq$500 kpc) may be more readily explained if this population of absorbers consists of pressure-confined tidal debris that was built up in small groups and clusters of galaxies over a Hubble time. They show that the space-density and cross-section of tidal tails in groups of galaxies are large enough that they could constitute a major source of the low redshift Ly$\alpha$ absorption features that are associated with galaxies. The space-density of groups within 10 Mpc of our galaxy is 2.4 $\times 10^{-3}$ Mpc$^{-3}$, which is close to the $\sim 6 \times 10^{-3}$ Mpc$^{-3}$ space-density calculated for Ly$\alpha$ absorbers, assuming they have a 1 Mpc radius. Other observational constraints on the properties of Ly$\alpha$ absorber such as their velocity dispersion, correlation properties, dimensions and abundances, can be shown to be consistent with this hypothesis.

## 5. Future Plans

Clearly much more data are needed. A number of groups are working on this. I am personally involved in two large programs. (i) Further data on very low column density Ly$\alpha$ absorbers is being obtained by a collaboration led by J. Stocke using the HST GHRS with bright Seyfert galaxies as background sources. These are generally at very low redshifts, and the galaxy redshift information needed for absorber-galaxy correlations is either already available (CfA redshift survey etc.), or can be obtained efficiently with wide field fiber systems working to modest magnitude limits. (ii) Many more QSO LOS have already been observed by the HST FOS[3]. By combining photometry from the KPNO 2.1m and Burrel Schmidt with data from the literature[6], together with wide field fiber spectroscopy to R=18 with the LCO fiber system and HYDRA on the WIYN telescope, and spectroscopy complete to R=21.5 over a field with 3 arcmin radius using the CFHT MOS system we hope to increase by an order of magnitude the number of absorber-galaxy pairs available for study.

## 6. Acknowledgements

I would like to thank my collaborators, particularly Ray Weymann, Houjon Mo and Sidney van den Bergh who did a large part of the work described above.